\documentclass[runningheads]{llncs}
\usepackage{graphicx}

\usepackage{float}
\usepackage{array}

\makeatletter
\def\convertto#1#2{\strip@pt\dimexpr #2*65536/\number\dimexpr 1#1}
\makeatother

\begin{document}
\title{Human-to-Human Strokes Recordings for Tactile Apparent Motion}
%
%
\author{
	Basil Duvernoy\inst{1}\orcidID{0000-0001-5429-5594} \and
	Sarah McIntyre\inst{1}\orcidID{0000-0002-0544-6533}
}
\authorrunning{B. Duvernoy and S. McIntyre}
%
\institute{Center for Social and Affective Neuroscience, Department of Biomedical and Clinical Sciences, Linköping University, Linköping, 58183 Sweden\\
	\email{duvernoy@liu.se}\\
	\url{liu.se/en/research/csan}
}
\maketitle              
\begin{abstract}
The main objective of this study is to investigate whether one can use recordings of human-to-human touch, such as a caress, to improve tactile apparent motion interfaces to make them feel more natural. We report here preliminary recordings of natural and continuous human-to-human caresses. To do this, six accelerometers were positioned on the receiving hand next to the stimulated area while a finger gently stroked the skin. The results suggest that we are able to capture signals from real human caresses that can be compared to signals produced by apparent motion stimuli. This is encouraging for our plan to continue the study in the second stage, which consists of tuning vibrotactile actuators to reproduce a similar pattern of vibrational responses in the accelerometers. In this way, the actuators mimic human behavior.  \\

\end{abstract}

\section{Introduction}

At the beginning of the century, vision researchers showed interesting results using a natural approach to study movement that can reveal more information than more classic methods~\cite{Felsen2005,Maallo2022}. 
In touch, most studies on apparent tactile motion investigate the perceptual impact of this technique by varying the values of predefined parameters~\cite{Israr2011,Hachisu2018,Kwon2021}. Even though this approach is classic, it is likely that the methods used so far to create apparent motion are not optimal. We expect that it should be possible to achieve more natural-feeling stimuli for the tactile apparent motion if they are tuned with recordings of natural tactile motions. 
Similar to previous apparent motion studies, we plan in the near future to use a set of vibrotactile actuators activated in sequence to produce the apparent motion illusion. With in mind to tune the actuators' behavior in a more natural approach, we recorded the vibrations generated by human-generated caresses, which constitute our case study. To do so, we used six accelerometers placed on the receiving hand. We were primarily interested in the speed and intensity of the overall motion, as well as the content of the frequency domain.

\section{Materials and Method}
\begin{figure}[H]
	\includegraphics[width=\textwidth]{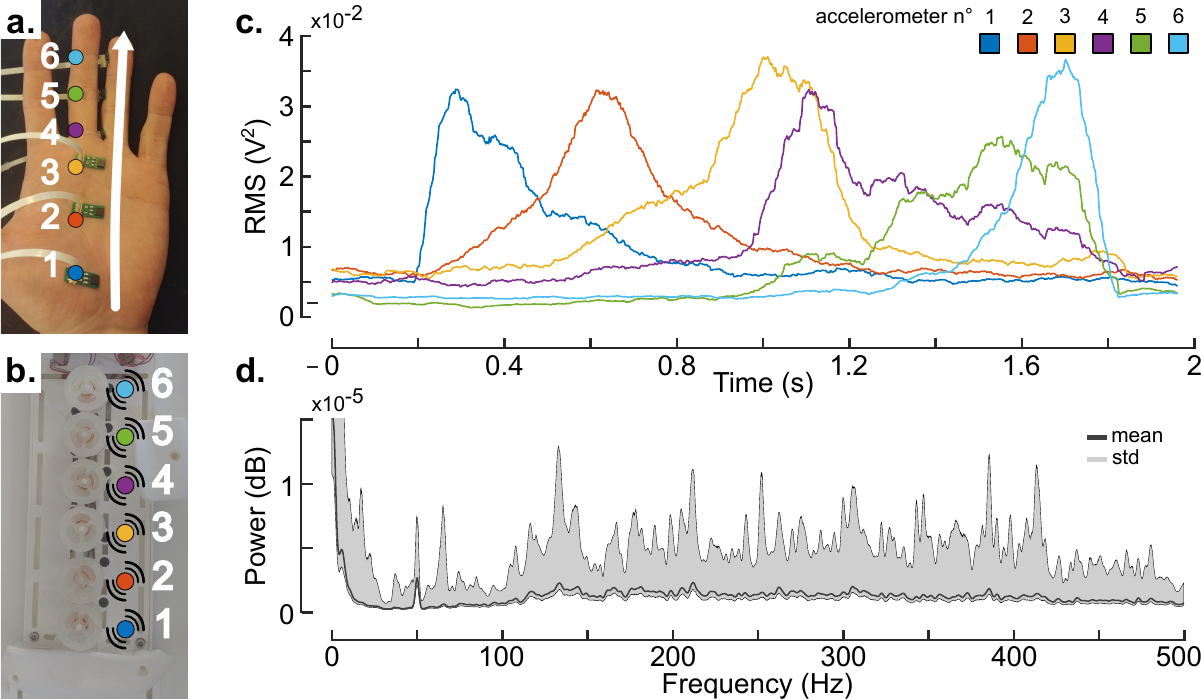}
	\caption{Natural observation: \textbf{a.} Accelerometers' locations and path of the stimuli; \textbf{b.}; apparent motion display for future work; \textbf{c.} Example of root mean square for each accelerometer during a single stroke from a finger; \textbf{d.} Overall mean and standard deviation of the power spectrum from the full data set.} \label{fig:results}
\end{figure}

\subsection{Apparatus}
Partly shown in figure~\ref{fig:results}a, the recording apparatus is composed of six 3-axis accelerometers (ADXL335 rev.B), three ADC (MCP3008), and a microcontroller (Teensy 4.1). The overall system is optimised to collect data up to 500~Hz at a sampling rate of 2~kHz.

\subsection{Procedure}
The tactile interaction involved the application of gentle stroking motions to a "receiver" hand using a single "transmitter" finger. The transmitter was required to stroke using her/his fingertip the receiving left hand from the bottom of the palm to the fingertip of the fourth finger (or vice versa). This location was chosen as the surface of the skin is relatively flat, which is required for the planned apparent motion device. The glabrous skin is also a sensitive area, where tactile cues can be more easily discriminated than other body places, such as the forearm. Therefore, the accelerometers were arranged linearly along the stimulated area, spaced at regular intervals of approximately 20~mm.

\section{Results}
Our preliminary result is based on recordings of 40 caresses. 
Figure~\ref{fig:results}c shows the root mean square (RMS) of the detrended recorded acceleration for each accelerometer using a 200~ms window. This highlights the information shared between neighboring accelerometers. A single caress is shown here, as the duration of the strokes was unsupervised and more processing is needed to compare the inter-accelerometers timings between stimuli. 
Figure~\ref{fig:results}d reports the mean and standard deviation of the power spectrum (PS) for each frequency in the 0-500~Hz range. This information has been obtained by applying a fast Fourier transform to the data from each accelerometer of each recording using a sliding Kaiser window of 128-samples on each sample; the resulting spectra were averaged and the standard deviation was extracted.

\section{Discussion}
RMS shows a similar pattern along accelerometer data: intensity increases to a peak, before decreasing. Signal overlaps can be explained by the propagation of waves on the skin, an important phenomenon for the future control of actuators. Planned additional analyses will reveal important information about the uniformity of stroking velocity and duration, which will inform apparent motion stimulus parameter optimisation. PS analysis shows that using one frequency to emulate apparent tactile motion may not be the best option. As previous studies of tactile apparent motion often used only one frequency, these findings may indicate a good way to provide more convincing tactile apparent motion. To achieve this, it is planned to use a bespoke vibrotactile display (see fig.\ref{fig:results}b) capable of delivering vibrations up to 500~Hz (based on this actuator design~\cite{Duvernoy2018}).

%
%
%
%

\end{document}